\documentclass[apj]{emulateapj}

\usepackage{natbib}

\usepackage{graphicx}
\usepackage[percent]{overpic}
\usepackage{lipsum}
\usepackage{multirow}
\usepackage{amssymb}
\usepackage{color}
\usepackage{rotating}
\usepackage{mwe,tikz}
\usepackage[percent]{overpic}
\usepackage{mathtools}
\usepackage{physics}
\usepackage{amsmath}
\usepackage[utf8x]{inputenc}
\usepackage{hyperref}
\usepackage{wasysym}
\usepackage{soul}

\newcommand{\msun}{${\rm M_{\sun}}$}
\def\ltsima{$\; \buildrel < \over \sim \;$}
\def\simlt{\lower.5ex\hbox{\ltsima}}
\def\gtsima{$\; \buildrel > \over \sim \;$}
\def\simgt{\lower.5ex\hbox{\gtsima}}
\def\kms{{\rm\,km\,s^{-1}}}
\def\pc{{\rm\,pc}}
\def\kpc{{\rm\,kpc}}

\def\msun{{\rm\,M_\odot}}


\def\deg{^\circ}

\def\Gyr{{\rm\,Gyr}}

\def\ltsima{$\; \buildrel < \over \sim \;$}
\def\gtsima{$\; \buildrel > \over \sim \;$}

\def\Khyatia#1{{\color{black}\bf #1}}  

\shorttitle{Origin of Globular Cluster Streams}
\shortauthors{Malhan et al.}

\begin{document}

\title{Butterfly in a Cocoon,\\
Understanding the origin and morphology of Globular Cluster Streams:\\
The case of GD-1}

\email{khyati.malhan@fysik.su.se}

\author{Khyati Malhan\altaffilmark{1}, Rodrigo A. Ibata\altaffilmark{2}, Raymond G. Carlberg\altaffilmark{3}, Monica Valluri\altaffilmark{4} and Katherine Freese\altaffilmark{1,5,6}}

\altaffiltext{1}{The  Oskar  Klein  Centre  for  Cosmoparticle  Physics,  Department  of Physics,  Stockholm  University,  AlbaNova,  10691  Stockholm,  Sweden}
\altaffiltext{2}{Universit\'e de Strasbourg, CNRS, Observatoire Astronomique de Strasbourg, UMR 7550, F-67000 Strasbourg, France}
\altaffiltext{3}{Department of Astronomy \& Astrophysics, University of Toronto, Toronto, ON M5S 3H4, Canada}
\altaffiltext{4}{Department of Astronomy, University of Michigan, Ann Arbor, MI, 48104, USA}
\altaffiltext{5}{The Nordic Institute for Theoretical Physics (NORDITA), Roslagstullsbacken 23, 106 91 Stockholm, Sweden}
\altaffiltext{6}{Leinweber Center for Theoretical Physics, Department of Physics, University of Michigan, Ann Arbor, MI 48109, USA}

\hspace*{\fill} NORDITA-2019-022, LCTP-19-06

\begin{abstract}
Tidally disrupted globular cluster streams are usually observed, and therefore perceived, as narrow, linear and one-dimensional structures in the 6D phase-space. Here we show that the GD-1 stellar stream ($\approx 30\pc$ wide), which is the tidal debris of a disrupted globular cluster, possesses a secondary diffuse and extended stellar component ($\approx 100\pc$ wide) around it, detected at $>5\sigma$ confidence level. Similar morphological properties are seen in synthetic streams that are produced from star clusters that are formed within dark matter sub-halos and then accrete onto a massive host galaxy. This lends credence to the idea that the progenitor of the highly retrograde GD-1 stream was originally formed outside of the Milky Way in a now defunct dark satellite galaxy. We deem that in future studies, this newly found {\it cocoon} component may serve as a structural hallmark to distinguish between the in-situ and ex-situ (accreted) formed globular cluster streams.
\end{abstract}


\section{Introduction}\label{sec:Introduction}

It is generally believed that large galaxies, like the Milky Way, underwent an initial {\it in-situ} formation phase, that was followed by {\it ex-situ} mass growth of the halo via merging and accretion of protogalactic  fragments \citep{Searle1978, Freeman2002}. This suggests that today’s galaxies should contain contributions from both {\it in-situ} and {\it ex-situ} formed tracer components, such as globular clusters (GCs), depending on the galaxy's assembly history. While, for the Milky Way, it is often assumed that the metal-rich GCs are associated with the {\it in-situ} phase of galaxy formation, and that metal-poor GCs are all accreted \citep{Forbes_Globular_Cluster_Review_2018}, this distinction is much harder to discern observationally, especially for the GC population in the stellar halo that shows a wide spread in metallicity \citep{Carollo2007metallicityhalo, Helmi2008_haloreview}. Yet this distinction is important to tightly constrain models of galaxy formation and evolution in a cosmological context. It is also relevant for dark matter studies, as accreted stellar objects tend to remain embodied within the dark envelope coming from the merging sub-halo progenitor; unlike stellar structures born within the Milky Way that are generally expected to be devoid of such dark sheaths.

The best example of GCs that are a result of accretion in the Milky Way is currently provided by those that are associated with the Sagittarius dwarf galaxy \citep{Bellazzini2003} and the LMC \citep{Wagner-Kaiser2017}. We know this to a certainty because we are witnessing the on-going merging process of the parent satellites. However, the GCs and the baryonic content of the  ancient accreted satellites, that were deposited into our Galactic halo $\gtrsim 5$ -- $6\Gyr$ ago, do not specifically feature any characteristic observable clues conveying the historical tale of their accretion. This is simply because such mergers have been gradually stripped of their dark matter envelopes and their observable stars have been mixed into the halo, leaving hardly any relic traces of their earlier existence. Therefore, one of the other approaches to distinguish accreted GCs is to instead survey the effects imprinted on their internal structure and kinematics (such as studying the evolution of their half mass radii or measuring their velocity anisotropies), which depends on the strength of the underlying tidal field. The idea is that GCs born in the massive Milky Way galaxy would be subjected to a different force field during their lifetime, than the ones that originated in dwarf galaxies and were later accreted. However, simulations show that post-accretion, the clusters adapt to the new tidal environment of their host galaxy, losing any signature of their original environment in a few relaxation times \citep{Meghan2016,Bianchini2017}. 

Here we aim to examine if narrow stellar streams, that are remnants of GCs, can be useful in addressing this problem and potentially dis-entangling the two different types of GC/stellar populations present in the halo.

\begin{figure*}
\begin{center}
\includegraphics[width=0.85\hsize]{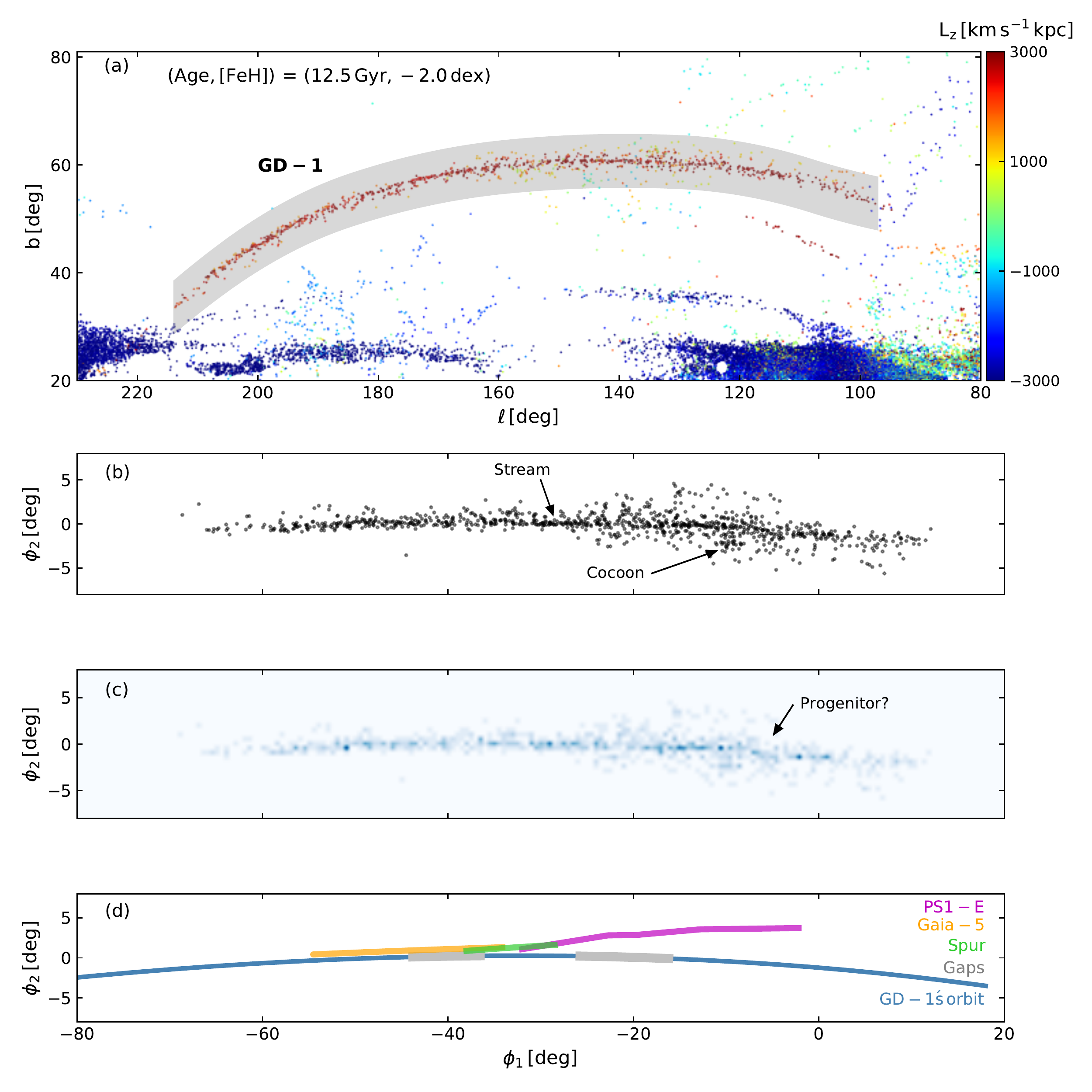}
\vspace{-0.5cm}
\end{center}
\caption{The cocoon around GD-1. (a) \texttt{STREAMFINDER}'s density map that was obtained from Gaia DR2 as described in Section \ref{sec:Diagnosing}, shown here in Galactic coordinates. The stars are color coded according to their z-component of angular momentum as calculated by the \texttt{STREAMFINDER}. The gray band is the $10\deg$ wide data selection that we made to isolate GD-1 stars of interest. (b) Stars corresponding to the data selection, shown here in a rotated coordinate system (introduced by \citealt{Koposov2010}) that aligns along the GD-1 stream. The plot reveals an extended stellar component, much like a {\it cocoon}, surrounding ``thin'' GD-1. (c) The number density plot corresponding to stars shown in (b) with $0.5\deg$ pixel resolution. (d) The approximate trajectories of the GD-1 stream, PS1-E stream, Gaia-5 stream and the spur along with the locations of the gaps are provided for comparison with our density map. These trajectories were produced by approximately curve fitting the on-sky extents of the respective structures as observed in previous studies \citep{Bernard2016, WhelanBonacaGD12018,Bonaca_spur_2018, Malhan_Ghostly_2018}.}
\label{fig:Fig_1}
\end{figure*}

A GC that undergoes tidal disruption due to its interaction with a massive host forms a long and thin stellar stream. Typically, their physical widths are comparable to the tidal radii of GC systems (a few tens of pc), and are therefore often observed to be quite narrow, linear and one-dimensional structures in the 6D phase-space, devoid of any extended structural component. Here we revisit one of the GC streams of the Milky Way, the GD-1 stream \citep{GrillmairGD12006}, and examine its structure to check if it possesses any morphological signature which can be useful in providing insights into the origin and the embryonic association of its progenitor system with an ancient satellite merger. 

This paper is arranged as follows. In Section~\ref{sec:Diagnosing} we present our diagnoses of the GD-1 structure, where we discover a secondary extended and diffuse stellar component around the thin stream at $>5\sigma$ confidence level. In Section~\ref{sec:Checking} we show that the detected component is a genuine feature and not an artifact of underlying background contaminants. In order to interpret our findings, we further analyze a set of synthetic streams produced in a cosmological simulations in  Section~\ref{sec:Simulations}, which ultimately assists us both in making a strong case for this detection and in understanding the origin of this secondary stream feature. We finally present our conclusion and discuss the prospects of our study in Section~\ref{sec:Discussion}.

\section{Diagnosing the GD-1 stream}\label{sec:Diagnosing}

The GD-1 stellar stream is observed as a long ($\sim 80\deg$, \citealt{WhelanBonacaGD12018}) and exceptionally narrow structure in the sky (angular width of $\sim 0.5\deg$), and was previously measured to have a physical width of $\sim 20\pc$ \citep{Koposov2010}. Its velocity dispersion in the direction tangential to the line of sight has been measured to be $<2.3\kms$ $(95\%$ conf.), and it is an extremely metal poor structure  $\rm{([Fe/H]}=-2.24\pm0.21)$ \citep{Malhan2018PotentialGD1}. The color-magnitude diagram (CMD) of GD-1 is similar to that of the M13 GC \citep{GrillmairGD12006} and corresponds to a stellar population of age $10$--$12\Gyr$. While the location of the progenitor of GD-1 is currently unknown \citep{Boer2018, Malhan_2018_PS1, WhelanBonacaGD12018, Webb2018GD1model}, and indeed it may have been completely disrupted long ago, these stream properties strongly imply that GD-1 is the remnant of a GC. 

We first detect the GD-1 stream structure of interest, using our \texttt{STREAMFINDER} algorithm \citep{Malhan2018_SF, Malhan_Ghostly_2018, Ibata_Phlegethon_2018,Ibata_Norse_streams2019}. The required stellar stream density map was obtained from processing the Gaia DR2 dataset, after adopting a Single Stellar population (SSP) template model of ${\rm (Age,[Fe/H])} = (12.5\Gyr, -2.0)$ from the PARSEC stellar tracks library \citep{Parsec_isochrones2012}. In order to detect the GD-1 stream in particular, the algorithm was made to process only those stars that lie in the region $80\deg < \ell < 230\deg$ and $b>20\deg$, and with Heliocentric distances in the range 1 to $20\kpc$. All other algorithm parameters are identical to those described in \cite{Ibata_Norse_streams2019}. The corresponding stream map is shown in Figure \ref{fig:Fig_1}a, where all the sources have a stream-detection significance of $>6\sigma$. This detection statistic means that at the position of each of these stars, the algorithm finds that there is a $>6\sigma$ significance for there to be a stream-like structure (with Gaussian width of $100\pc$, and $\pm 10\deg$ long) passing through the location of each of the stars, and with proper motion consistent with that of the stars.

The algorithm successfully detects the $80\deg$ long GD-1 stream, as well as numerous other tentative stream structures, which will not be discussed further in this contribution. To isolate the GD-1 stars of interest we make a $10\deg$ wide selection around the stream track (gray band in Figure \ref{fig:Fig_1}a). The corresponding selected stars are shown in panel (b) of Figure \ref{fig:Fig_1} in the rotated coordinates of GD-1 (the transformation from the galactic coordinate system to this new rotated frame was made using the conversion matrix provided in \citealt{Koposov2010}). This map triggered our initial suspicion that the GD-1 stream, that is usually conceived as a physically thin structure, could possibly be part of an extended and diffuse stellar component (dubbed {\it ``cocoon''} in the Figure). The dense stellar structure lying along $\phi_2 \sim 0\deg$ is the GD-1 stream. This contribution is devoted to the study of this extended component and its origin. Figure \ref{fig:Fig_1}c shows the star number density map, corresponding to Figure \ref{fig:Fig_1}b, obtained using a pixel size of $0.5\deg$. 

Recent studies have revealed several interesting structures in the region of the sky surrounding GD-1: these include the neighbouring streams (PS1-E, \citealt{Bernard2016}, \citealt{Malhan_2018_PS1} and Gaia-5, \citealt{Malhan_Ghostly_2018}), the detection of possible gaps in GD-1 \citep{WhelanBonacaGD12018}, and a ``spur'' structure \citep{Bonaca_spur_2018}. 
These features are sketched schematically in Figure \ref{fig:Fig_1}d. The number density map in Figure \ref{fig:Fig_1}c reveals some of these known features, such as the gaps at $\phi_1 \sim -40\deg$ and $-20\deg$, and also part of the PS1-E stream. The map also reveals a possible location for GD-1's elusive missing progenitor at $[\phi_1, \phi_2] \approx [-6\deg, -0.8\deg]$, which we tentatively propose is associated with the ``kink'' feature (characteristic of stars emerging from the Lagrange points of the progenitor of a tidal stream) that is visible at this location. 

In our analysis below, we ensure that the identified signal corresponding to the conspicuous extended stellar component (the {\it cocoon} in Figure \ref{fig:Fig_1}b) does not arise due to the presence of these surrounding stellar structures.

%
\subsection{Data}\label{subsec:Data}

The  \texttt{STREAMFINDER} algorithm was made to process only those stars for which Gaia's extinction corrected magnitude $G_0<19.5$. The magnitude cut was chosen so as to avoid edge-effects stemming from the faint limit of our Gaussian Mixture Model at $G_0= 20.0$ (see \citealt{Ibata_Norse_streams2019} for technical details). Also, the number of sources rise towards the fainter magnitude limit, that makes the processing time of the algorithm computationally expensive. We therefore made a fresh sample selection directly from the Gaia DR2 dataset in order to study the GD-1 and the accompanying feature of interest. 

We take the Gaia DR2 catalogue for the region of the sky ranging between $(\phi_1,\phi_2)=([-60\deg, 20\deg], [-15\deg, 15\deg])$, consistent with the on-sky extent of the GD-1 stream, and first correct it for extinction using  the \cite{Schlafly2011} corrections to the \cite{Schlegel1998} extinction  maps,  assuming  the  extinction ratios $A_G/A_V=  0.85926,A_{G_{\rm BP}}/A_V=  1.06794, A_{G_{\rm RP}}/A_V=  0.65199$,  as  listed  on  the  web  interface  to the  PARSEC  isochrones \citep{Parsec_isochrones2012}. Henceforth all magnitudes and colors refer to these extinction-corrected values. For this data selection, we also make cuts in Gaia's color $(0.40<[G_{\rm BP}-G_{\rm RP}]_0<1.60)$ and the magnitude $(G_0<20)$ space. The color window ensures the inclusion of the GD-1 like stellar populations, discarding most of the disk contaminants. The chosen magnitude limit mitigates against the effect of completeness variations due to inhomogeneous extinction. We set constraints also in the proper motion $(\sqrt{{\mu^{*}_{\alpha}}^2 + \mu_{\delta}^2}>2$, where $\mu^{*}_{\alpha} \equiv \mu_{\alpha}cos \delta$) and the parallax $(\overline{\mathbb{\omega}}<0.33 \rm{\,mas\,at\,3\sigma\,level})$ selection in order to include only those stars that inhabit a similar phase-space region as GD-1. Henceforth, for convenience, we will drop the asterix superscript from $\mu^{*}_{\alpha}$. This sample of stars was then cross-matched with the spectroscopic surveys of SEGUE (SDSS DR10, \citealt{SEGUE_SDSS2009}) and LAMOST DR4 \citep{LAMOST2012} datasets in order to acquire their line-of-sight velocities $(v_{\rm los})$ that are missing in Gaia DR2\footnote{Gaia DR2 measures $v_{\rm los}$ only for the stars with ${G}\simlt13$.}. From these cross-matches we obtained their $v_{\rm los}$ and ${\rm [Fe/H]}$ measurements. We then imposed a very conservative metallicity cut in the cross-matched catalogue, selecting stars with ${\rm [Fe/H]} < -1.5$ (consistent with GD-1's metallicity, following \citealt{Malhan2018PotentialGD1}) so as to retain a maximal population of GD-1-like stars and reject most of the contaminants. We retain this filtered dataset and refer to this as sample-1.

\begin{figure}
\begin{center}
\includegraphics[width=1.06\hsize]{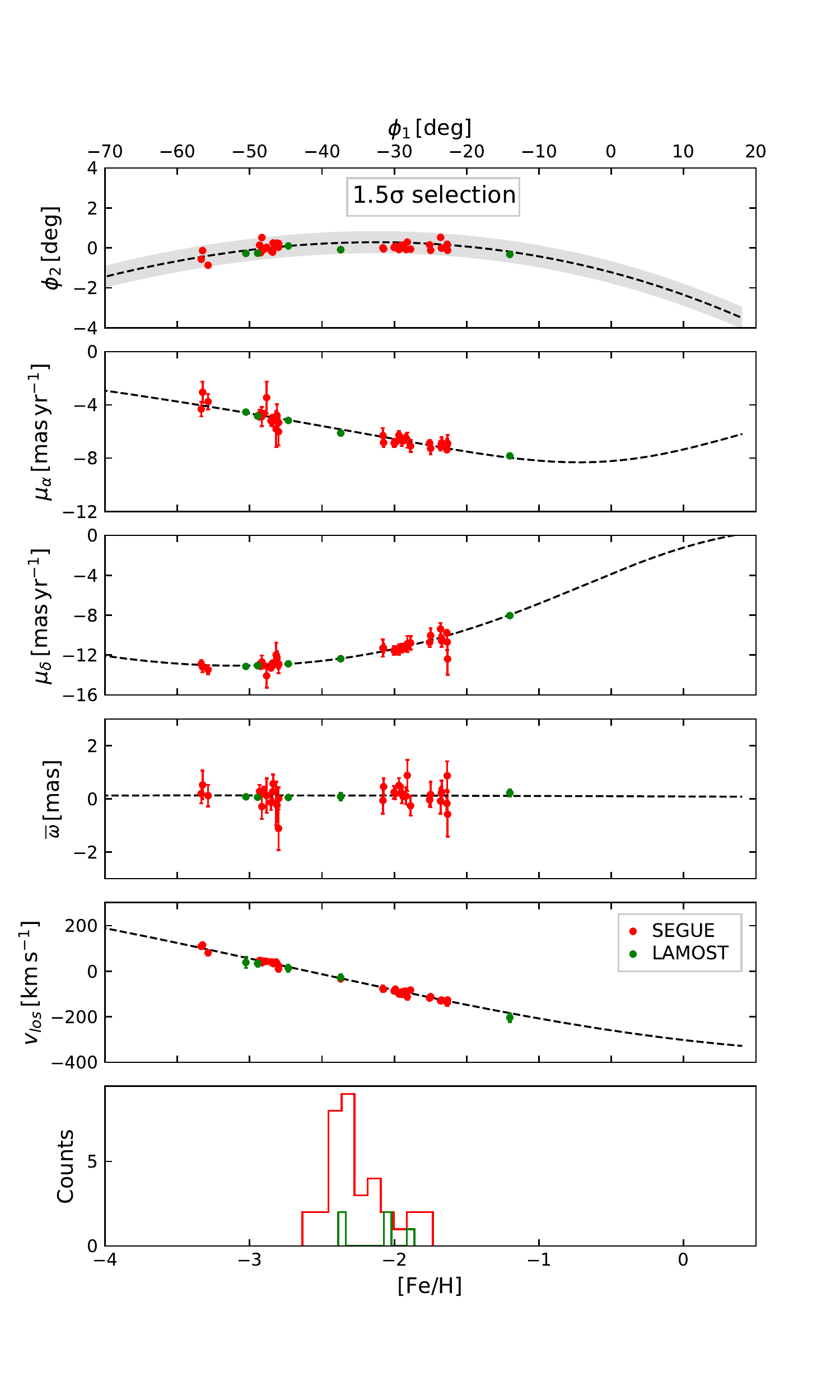}
\vspace{-1.5cm}
\end{center}
\caption{The phase-space-metallicity distribution of GD-1 stars, based on selection within a $1.5\sigma$ threshold of the adopted GD-1 orbital model. From top to bottom the panels show the angle $\phi_2$ in the standard GD-1 coordinate system, proper motion $\mu_{\alpha}$ and $\mu_{\delta}$, the parallax $\overline{\mathbb{\omega}}$,  and the heliocentric line of sight velocity $v_{\rm los}$ of the stream stars, as a function of $\phi_1$. The lowest panel shows the metallicity distribution of the stars. The orbital model of GD-1 is shown as a dashed curve while the data points are shown in colour. The light gray band along the model highlights the effective $1.5\sigma$ region around the orbit. This stringent selection yields $40$ stars that are shown here.}
\label{fig:Fig_2}
\end{figure}

\begin{figure}
\begin{center}
\includegraphics[width=\hsize]{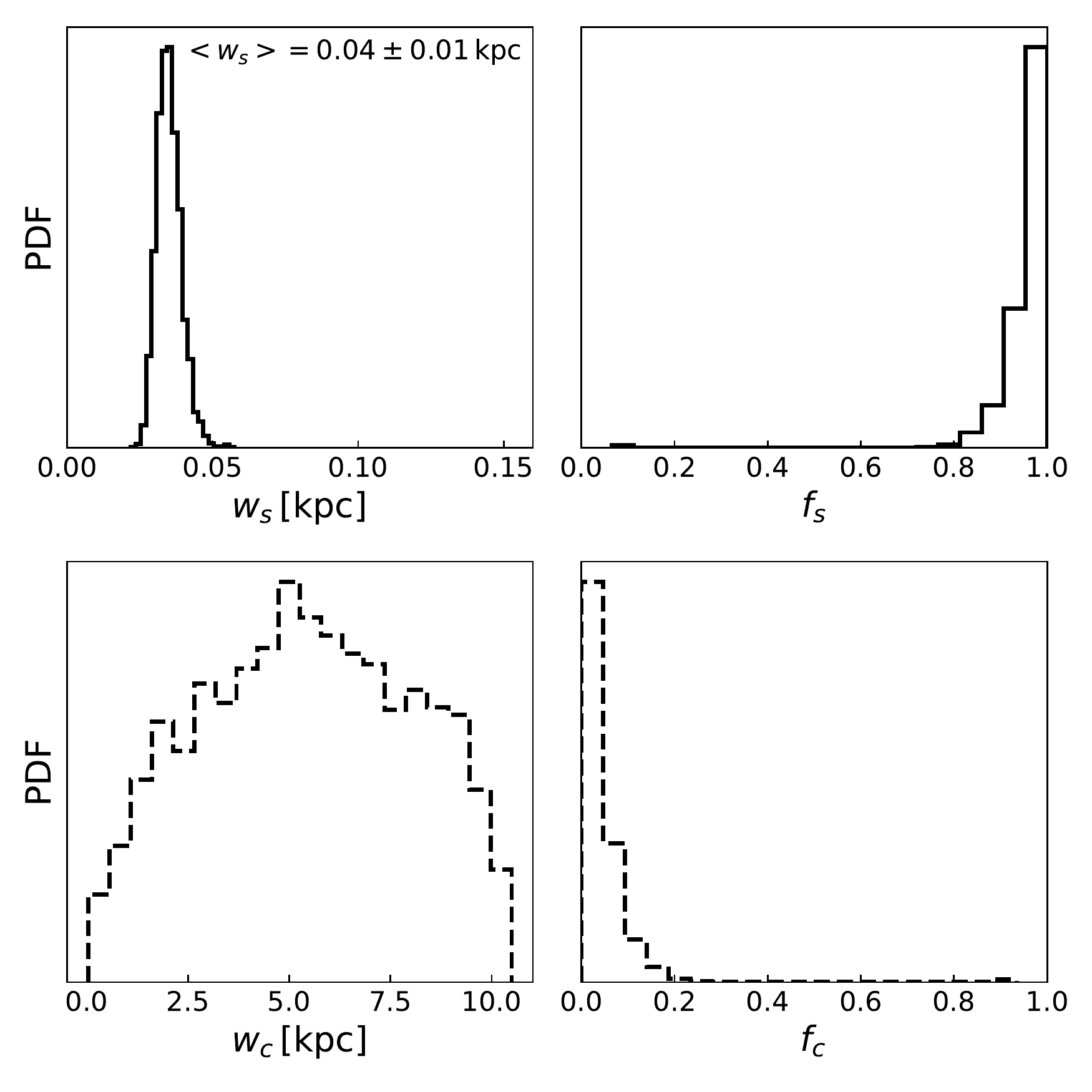}
\vspace{-0.5cm}
\end{center}
\caption{Probability distribution function for the 4-parameter Gaussian model (described in Equation \ref{eq:test1}) corresponding to the data points shown in Figure~\ref{fig:Fig_2}. The top plots here represent the distribution of $({w_s, f_s})$, while the bottom plots show the same but for $({w_c, f_c})$. Since ${f_s \to 1 (f_c \to 0)}$, this suggests that the $1.5\sigma$ selection does not show significant evidence of any secondary features, and consists of only a uni-modal structural component, GD-1 itself, that is $40\pm10\pc$ wide. This is not surprising, as the stringent sample selection (made in the multi-dimensional volume of phase-space and photometry information) can barely accommodate any additional structural feature.}
\label{fig:Fig_3}
\end{figure}
\begin{figure*}
\begin{center}
\vspace{-0.30cm}
\includegraphics[width=0.8\hsize]{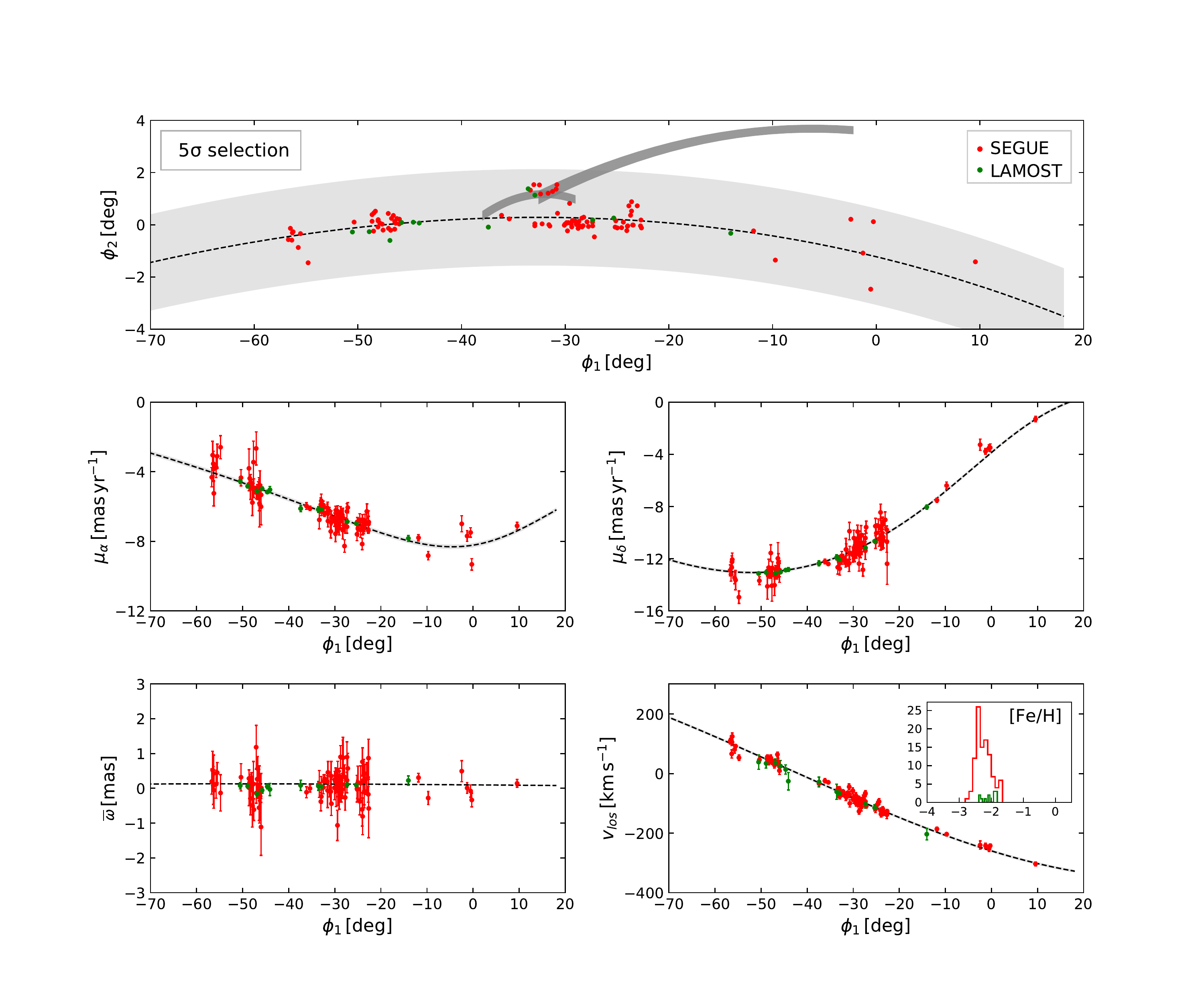}
\vspace{-1.0cm}
\end{center}
\caption{ Same as Figure \ref{fig:Fig_2}, but for a comparatively relaxed sample selection with a $5\sigma$ selection threshold. The sample contains  $116$ stars which are shown here. The light gray band along the GD-1 track highlights the effective $5\sigma$ region around the orbit, while the dark gray curves in the top panel highlight the locations and extents of the spur \citep{Bonaca_spur_2018} and the PS1-E stream \citep{Bernard2016}.}
\label{fig:Fig_4}
\end{figure*}
\begin{figure}
\begin{center}
\includegraphics[width=\hsize]{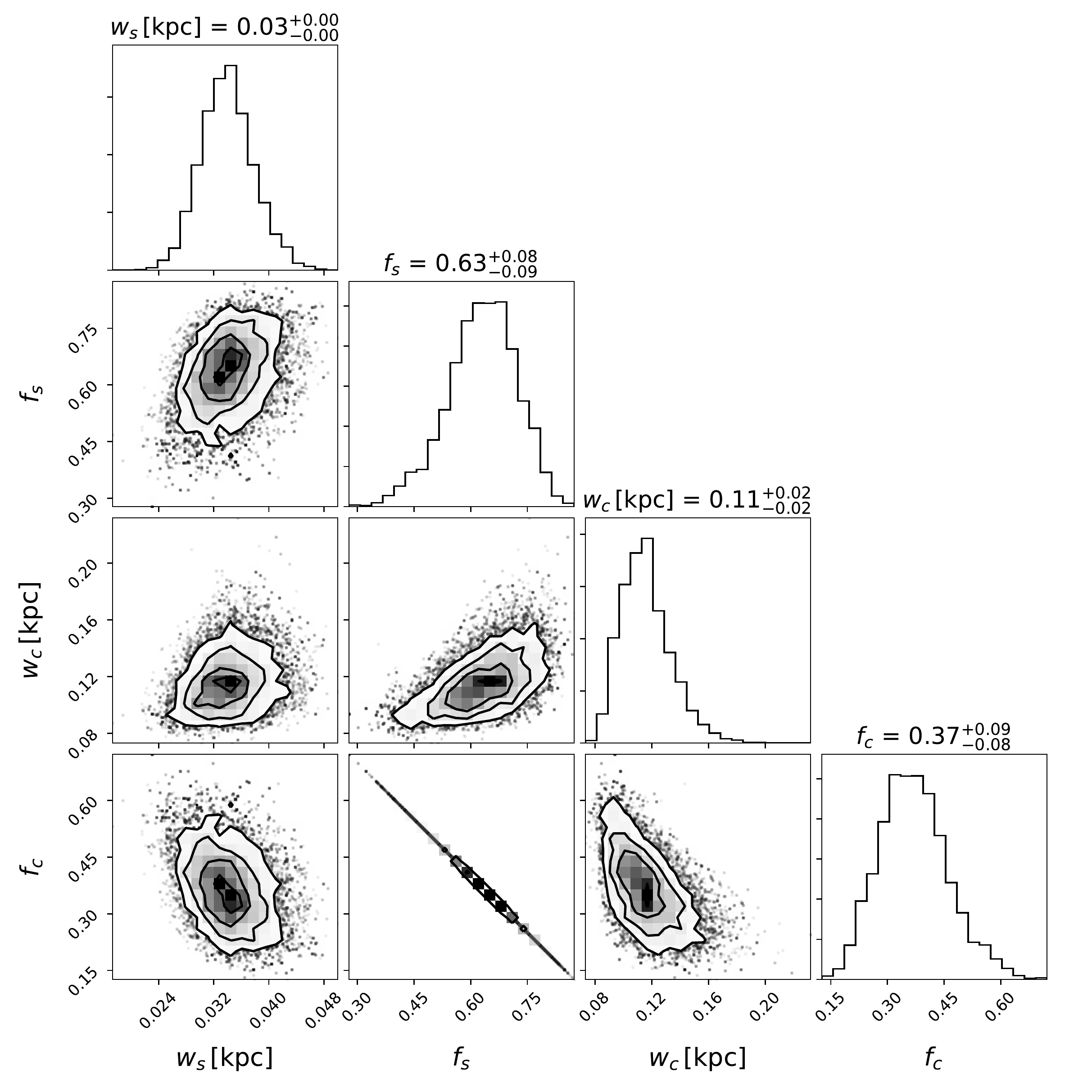}
\vspace{-0.8cm}
\end{center}
\caption{Probability distribution function for the 4-parameter Gaussian model (described in Equation \ref{eq:test1}) corresponding to the $5\sigma$ data selection threshold shown in Figure~\ref{fig:Fig_4}. In addition to identifying the GD-1 stream with $\langle {w_s} \rangle = 30\pc$ and $\langle {f_s} \rangle = 0.63$, we also detect an underlying secondary diffuse component with $\langle {w_c} \rangle = 110\pm20\pc$ and $\langle {f_c} \rangle = 0.37$. This means that  $37\% (\approx 43$ stars) of the sample belongs to this newly-identified structure. We associate this secondary structure with the GD-1 {\it cocoon}.}
\label{fig:Fig_5}
\end{figure}
\subsection{Anatomy of the GD-1 stream}\label{subsec:Anatomy}
 
In \cite{Malhan2018PotentialGD1} we performed an orbit-fitting procedure for the GD-1 stars, although to a more conservative star sample (in particular, the spatially-extended component was rejected), in order to constrain the gravitational potential of the Milky Way. One of the natural by-products of the study was the best-fit orbit model for the GD-1 stream. We use the same orbit model here for the purpose of selecting stars in sample-1.

First, we analyze the stars contained in a very narrow and restricted phase-space region around GD-1. This was done by sigma-clipping and selecting only those stars in sample-1 that lie within $1.5\sigma$ of the orbit model in the observed phase-space parameters $(\phi_1,\phi_2, \overline{\mathbb{\omega}},\mu_{\alpha}, \mu_{\delta}, v_{\rm los})$\footnote{While performing sigma clipping for parallaxes $\overline{\mathbb{\omega}}$, we accounted for the zero-point correction of the parallax measurements present in Gaia DR2 \citep{GaiaDR2_2018_astrometry}.}. Here, $\phi_1, \phi_2$ refer to the angles of the GD-1 coordinate system as mentioned previously, $\overline{\mathbb{\omega}},\mu_{\alpha}, \mu_{\delta}$ are the parallax and proper motion measurements that come from the Gaia data, and $v_{\rm los}$ refers to the heliocentric line of sight velocity obtained from our cross-matches with the SEGUE and LAMOST catalogues. This stringent selection ensures the rejection of any datum that is even remotely an outlier. Simultaneously, we also clip the data in Gaia's colour-magnitude $(G_0, [G_{\rm BP} - G_{\rm RP}]_0)$ space using the same SSP model that previously allowed GD-1's detection. This very fine selection yields a total of $40$ stars ($35$ from SEGUE and $5$ from LAMOST), that represent very high confidence GD-1 candidate members. These stars are shown in Figure \ref{fig:Fig_2}. 

We then test if this stringent sample selection of GD-1 members contains only a uni-modal stellar component (GD-1 itself), or accommodates any additional feature. This we investigate using a generative model which we take to be a two component Gaussian model; one component to account for the stream itself and the other for any possible secondary feature. The free parameters for our model were then provided as ${(w_s,f_s)}$ and ${(w_c,f_c)}$. Here ${w_s}$ and ${f_s}$  denote the stream's physical width and the fraction of stars in the data contained within the stream, and ${w_c,f_c}$ have the same meaning but for the secondary population. The likelihood function was expressed as:

\begin{equation}\label{eq:test1}
\begin{split}
\mathcal{L} &= \prod_{\rm d}\Big\{  \frac{f_s}{\sqrt{2\pi} w_s E(p,w_s)} {\exp} \Big[-\frac{1}{2}\Big(\frac{\phi^m_2 - \phi^d_2}{w_s}\Big)^2\Big]+ \\
& \frac{f_c}{\sqrt{2\pi} w_c E(p,w_c)}{\exp}\Big[-\frac{1}{2}\Big(\frac{\phi^m_2 - \phi^d_2}{w_c}\Big)^2\Big]\Big\} \Big|_{\phi_1}\\, 
\end{split}
\end{equation}
where ${f_c\equiv1-f_s}$. Here, $\phi^d_2$ is the observed position of a data point, and the corresponding orbit model value is given by $\phi^m_2$. Notice that since we are interested in measuring the physical dispersion of the structure(s) in the direction perpendicular to the GD-1's orbit, this calculation is made only in the position space for the $\phi_2$ coordinate. $E(p,w)$ is the error function which is included as a modification to the normalisation of the Gaussian function. This factor takes into account the sigma clipping procedure that we undertake for the data selection which abruptly truncates the data at a given $\sigma-$value. Here, it is defined as $E(p,w)=\mathrm{erf}(p/w \sqrt2)$, and in this case $p=1.5$. We then sample the posterior probability distribution function (PDF) with an MCMC algorithm. The resulting PDF is shown in Figure \ref{fig:Fig_3}. The PDFs are well behaved and suggest that the dataset under inspection contains only a single uni-modal structural component, since ${f_s \to 1}$,  corresponding to a physical width of $\langle {w_s} \rangle = 40\pm10\pc$. This means that with this (extremely narrow) sample selection our generative model only detects the thin GD-1 stream.

Next, we make a comparatively lenient selection from sample-1 by selecting all the stars in sample-1 that lie within a $5\sigma$ threshold with respect to the orbital track and the assumed SSP model. The selection yields a total of $116$ stars ($104$ from SEGUE and $12$ from LAMOST) and the corresponding stars are shown in Figure \ref{fig:Fig_4}. Once again, we fitted the same double Gaussian model via an MCMC process (in this case $p=5$). The PDFs were found to be well behaved (Figure \ref{fig:Fig_5}), albeit this time revealing a bi-modal distribution given the detection of a secondary extended component alongside GD-1. This secondary component was identified with a physical width of $\langle {w_c} \rangle =110\pm20\pc$, with $37\% \, {(f_c =0.37)}$  of the stars belonging to this newly-identified component ($\approx 43$ stars). We dub this component the {\it cocoon} of GD-1. The revised physical width of the narrow component is found to be $\langle {w_s} \rangle \sim 30\pc$.

We also realized that such a relaxed selection, although made in the multi-dimensional volume of phase-space and photometry information, may also gather stars from the structures lying in GD-1's neighbourhood (shown in Figure~\ref{fig:Fig_1}d). Therefore, in order to ascertain that this newly identified {\it cocoon} component is not a reflection of these surrounding localized structures, we repeat our previous analysis except this time masking the sky regions where GD-1's neighbouring structures lie. To mask the spur, for instance, we first approximate its on-sky trajectory from the map of \cite{Bonaca_spur_2018} and adopt its width to be $0.2\deg$, as per their study. We repeat the same for the PS1-E stream \citep{Bernard2016}, and set its width at $0.3\deg$. These models (or masked regions) are shown in Figure~\ref{fig:Fig_4} (top panel), and the stars lying in these regions were then removed. We do not mask the Gaia-5 stream region because, although it happens to be located close to GD-1 in the sky, Gaia-5's proper motions are significantly different from those of GD-1 \citep{Malhan_Ghostly_2018}, and hence it is not enclosed in the $5\sigma$ threshold selection that we make here. This masked selection, in contrast to the previous unmasked case, yields a total of $109$ stars ($98$ from SEGUE and $11$ from LAMOST). We again find a bi-modal distribution of stars, but with a slight variation in our parameter values. This time, the secondary component was identified with a physical width of $\langle {w_c} \rangle =130^{+30}_{-20}\pc$, with $22\% \, {(f_c =0.22)}$ of the stars belonging to this extended component ($\approx 24$ stars). The observed dip in the {\it cocoon}'s signal strength in this case is expected, as the $7$ stars that were now masked previously lay in the exterior region of GD-1, thereby contributing to the {\it cocoon} signal.

Having established the presence of a secondary extended stellar component, we then take the likelihood ratio test to assess the need for the presence of two populations rather than just one. For this, we used the unmasked (masked) data sample of $116\,(109)$ stars as above, and fitted them using the same double Gaussian model except this time forcing $f_c = 0$ and $w_c=1$ (the value of $w_c$ is unimportant given that $f_c$ is set to zero). The corresponding $p$-value with respect to the fully free model above is $p=1.46 \times 10^{-6} \, (p=1.18\times 10^{-7})$, indicating that the simpler model with a single uni-modal width distribution can be rejected with high confidence at the $5\sigma\,(5\sigma)$ level. Hence, the evidence for an additional component appears to be strong. Note that the estimated value of $w_c$ in our analysis is sensitive to the phase-space volume probed for the data selection (set to a $5\sigma$ threshold width here), nevertheless it can be perceived to be a lower bound. This means that in reality the detected {\it cocoon} structure may be spatially extended to even greater distances from GD-1, as was initially suggested by Figure~\ref{fig:Fig_1}b. This property is also revealed by the synthetic streams that are produced in cosmological simulations (we discuss this in Section~\ref{sec:Simulations} below).

\begin{figure*}
\begin{center}
\includegraphics[width=\hsize]{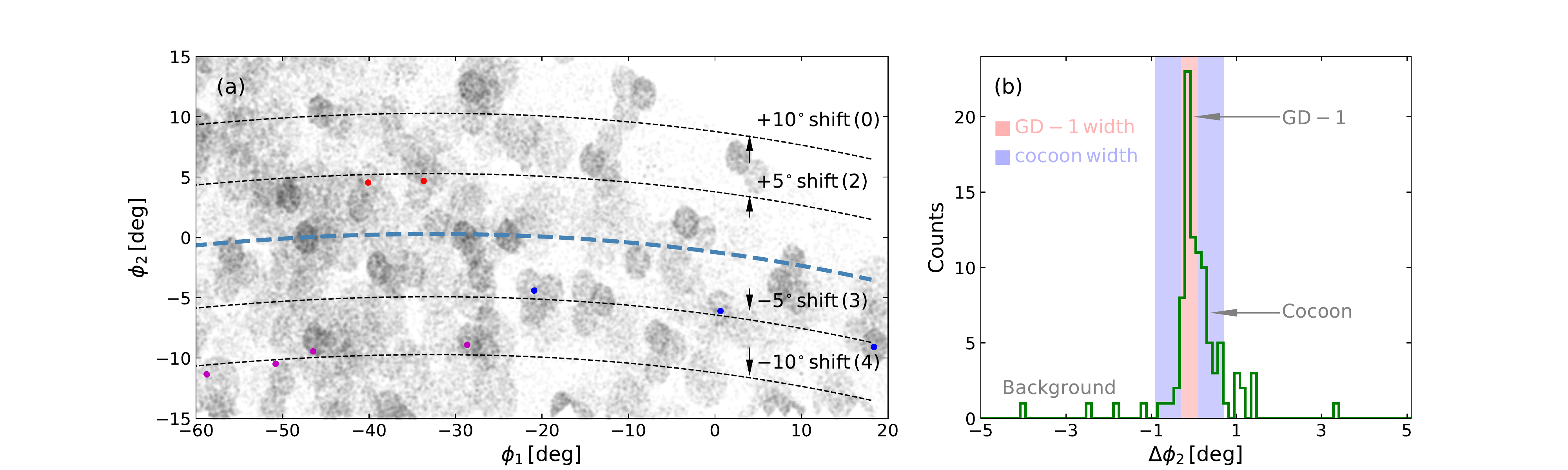}
\vspace{-0.5cm}
\end{center}
\caption{{\it Left panel:} Checking for contamination along GD-1's track. The gray background shows the field star number density of our sample-1 data as obtained in Section \ref{sec:Diagnosing} via cross-matching the Gaia DR2 catalog with the SEGUE and LAMOST datasets. This density map appears patchy due to the non-uniform sky coverage of these two spectroscopic surveys. GD-1's orbital track is shown in cyan. We shifted this track in the $\phi_2$ coordinate, using different values of $\phi_2$ as marked in the diagram, to test the level of field star contamination in the surrounding regions. A $5\sigma$ threshold for selection was used throughout. The number of contaminants found are reported in brackets, and also shown as color points, and are substantially smaller than the number found for the {\it cocoon} structure that we detect in this study. {\it Right panel:} Stellar density distribution, obtained by making a $3\sigma$ threshold selection in proper motions, parallaxes, photometry and line of sight velocity with respect to the original GD-1 orbital model. $\Delta \phi2$ refers to the angular difference between the datum and the GD-1 orbit. The \Khyatia{green} profile reveals a narrow peak at $\Delta \phi2 \sim -0.1\deg$ (the ``thin'' GD-1 stream) along with a broadened distribution (the {\it cocoon}) that extends out to $\sim 1\deg$ on either direction. The red and blue regions represent the estimated (on-sky) widths of the thin GD-1 stream and the extended component, respectively.}
\label{fig:Fig_6}
\end{figure*}
\section{Cocoon or Contamination?}\label{sec:Checking}

In order to ascertain the existence of this {\it cocoon} feature, it is important to ensure that it is not an artifact of the underlying background contamination that lies in the region of the sky containing GD-1. To examine this issue, and to quantify the contamination level present along GD-1's orbital track, we followed a tailor-made test that we now describe. 

We take GD-1's orbit model and shift it by $+10\deg$ in $\phi_2$, keeping its configuration fixed in the other phase-space dimensions, and count the number of stars that get selected by the orbit within $5\sigma$ from sample-1 in the same fashion as discussed before. This is shown in Figure \ref{fig:Fig_6}a. In this case, the selection around the orbit draws 0 contaminants. We repeat this procedure for $+5\deg, -5\deg, -10\deg$ shifts in $\phi_2$ from the $0\deg$ reference point, and respectively find that only 2, 3 and 4 contaminants are selected around the orbit. Thus, the observed level of {\it cocoon} members is unlikely to be due to random contaminants.

Next, in order to extract out the information on the spatial extent of the {\it cocoon} we do the following. Keeping GD-1's orbit model fixed at its original position, we make a $3\sigma$ threshold selection of the data points in proper motion, parallax, photometry and los velocity information space. For each selected datum, the angular difference $\Delta \phi_2$, between the GD-1's orbital model and the observed value, is calculated and the data is binned at that value. The corresponding stellar density distribution is shown in Figure \ref{fig:Fig_6}b. The narrow peak at $\phi_2 \sim -0.1\deg$ shows the presence of the ``thin'' component of the GD-1 stream, while the broadened distribution reveals the presence of the {\it cocoon} that extends out to $\sim 1 \deg$ in either direction.

\begin{figure*}
\begin{center}
\includegraphics[width=\hsize]{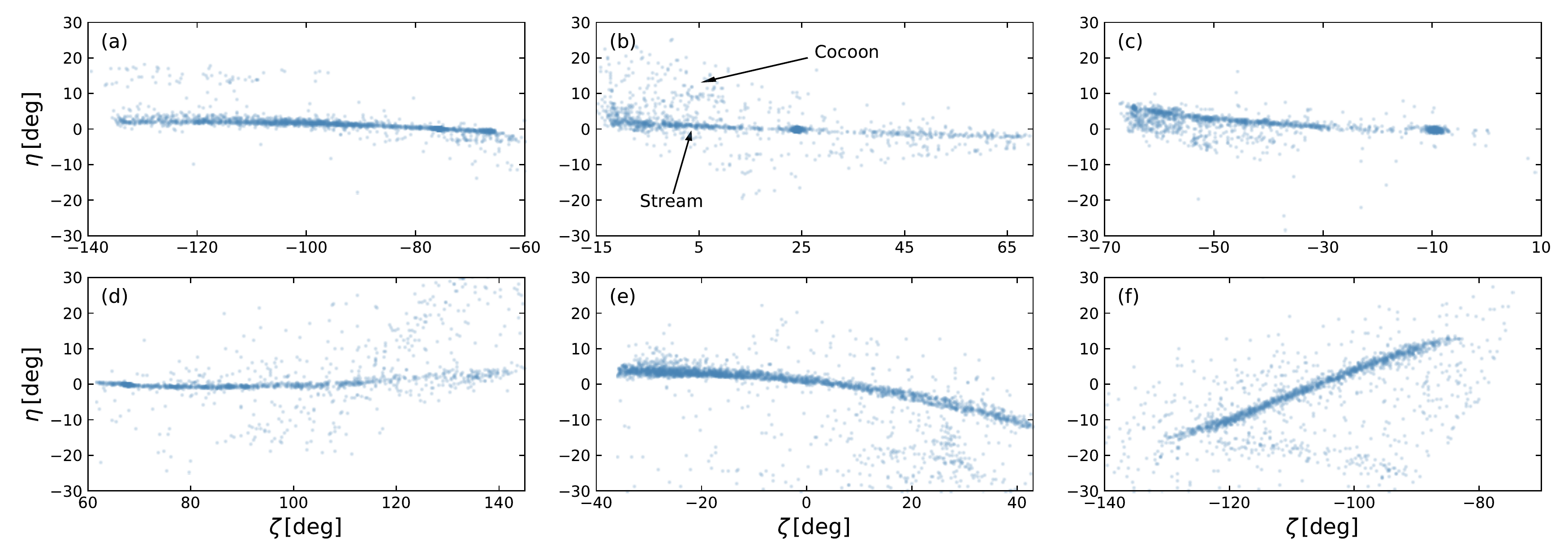}
\vspace{-0.6cm}
\end{center}
\caption{GC streams in a cosmological simulation. Stellar stream particles for some of the synthetic streams are shown in their rotated coordinate systems. Particles shown in each panel are member stars belonging to the same GC progenitor. Almost all the synthetic streams, of which only 6 are presented here, featured an additional distinctive diffuse stellar component (which we dub the {\it cocoon} in our analysis). This secondary feature, accompanying the thin GC stream, is likely a ubiquitous characteristic exhibited by all the streams that are remnants of the accreted GC streams that arrived within their own dwarf galaxy-mass dark matter sub-halos.}
\label{fig:Fig_7}
\end{figure*}
\begin{figure*}
\begin{center}
\includegraphics[width=0.21\hsize]{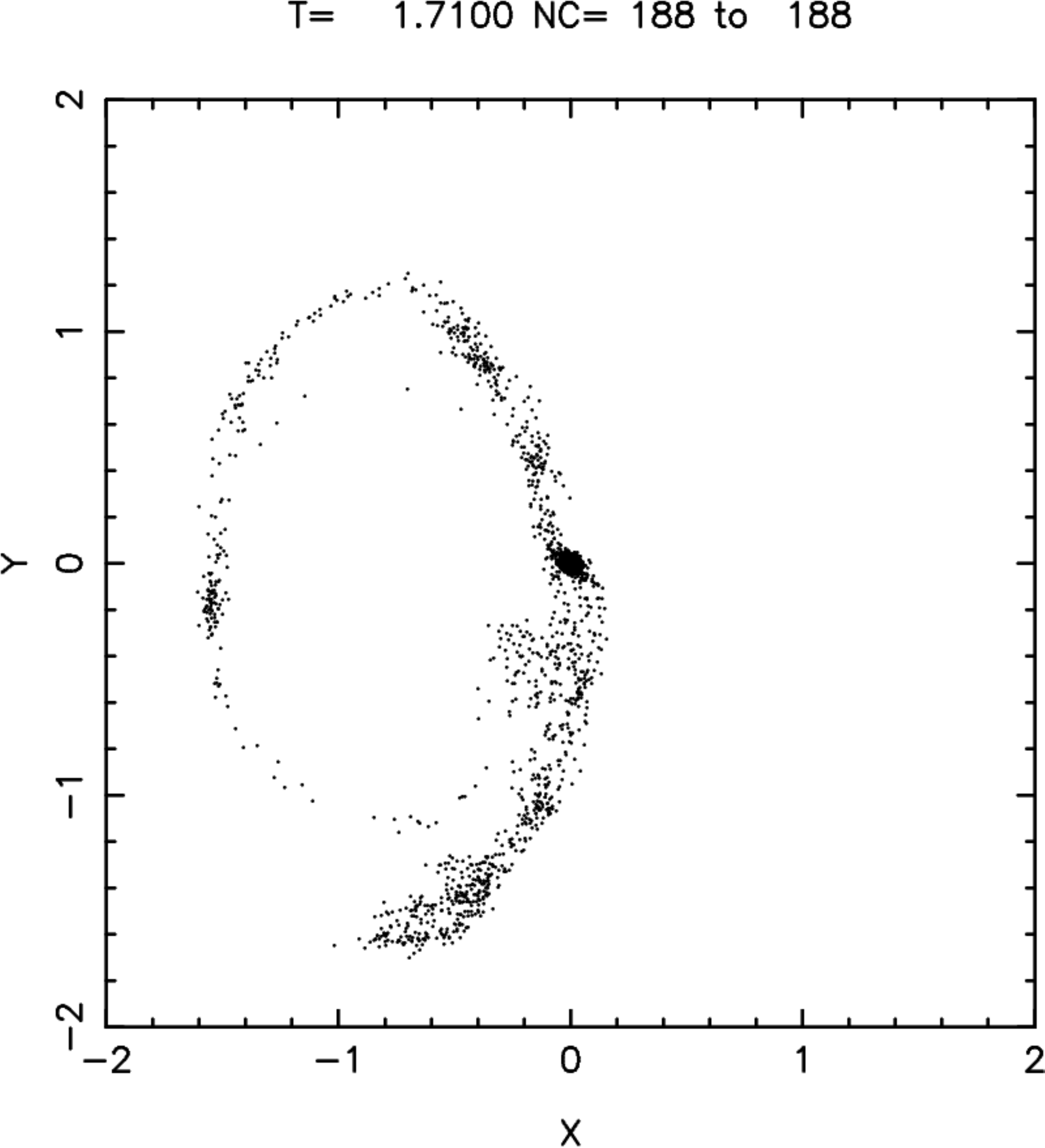}
\hspace{0.15cm}
\includegraphics[width=0.21\hsize]{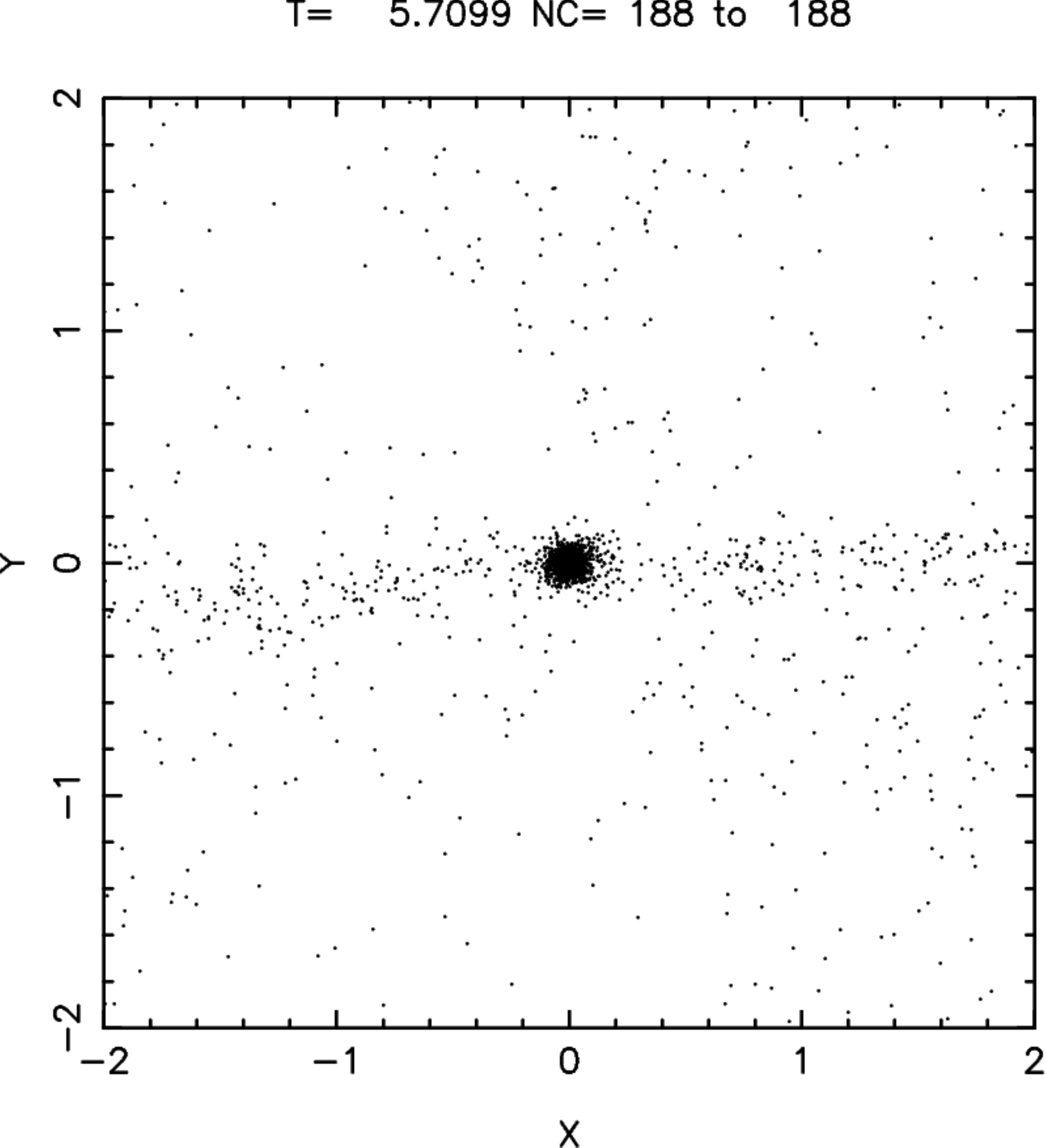}
\hspace{0.15cm}
\includegraphics[width=0.21\hsize]{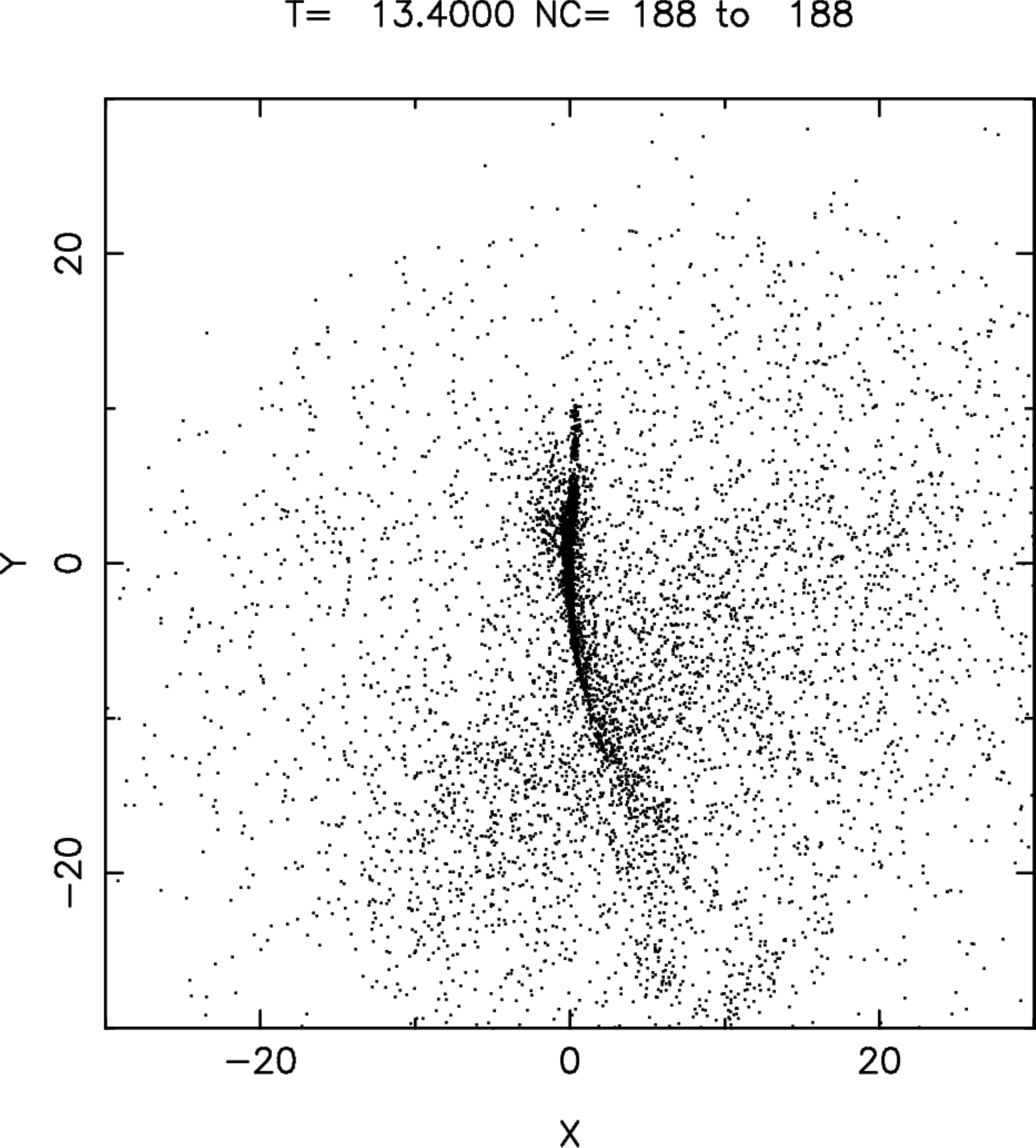}
\end{center}
\vspace{-0.2cm}
\caption{Dynamical evolution of GCs in the cosmological simulation (see Section~\ref{sec:Simulations} for details). Only the GC and its associated stellar particles are shown here.  The frames (from left to right, in kpc units) show the dynamical evolution and tidal disruption of a GC as it is accreted onto the main halo within its dark matter sub-halo. The X and Y coordinates are centred on the GC system. The {\it left panel} ($T\approx 1.7\Gyr$) shows the GC's pre-merging ``donut'' phase as the stars undergo tidal stripping within the sub-halo. The blob at the center is the progenitor GC in all panels. The {\it middle panel} ($T\approx 5.7\Gyr$) shows the initial stage of accretion of the GC, with its sub-halo, onto the main halo. The {\it right panel} ($T\approx 13.4\Gyr$, the scale of this panel is wider) shows the present epoch. The dark matter envelope is almost completely disrupted.  On the other hand, the GC ends up forming a thin and dense tidal stellar stream, and the stars that were spread out in the ``donut'' like shape end up creating a broad stream (the {\it cocoon}) surrounding the thin stream structure.}
\label{fig:Fig_8}
\end{figure*}
\begin{figure}
\begin{center}
\includegraphics[width=0.9\hsize]{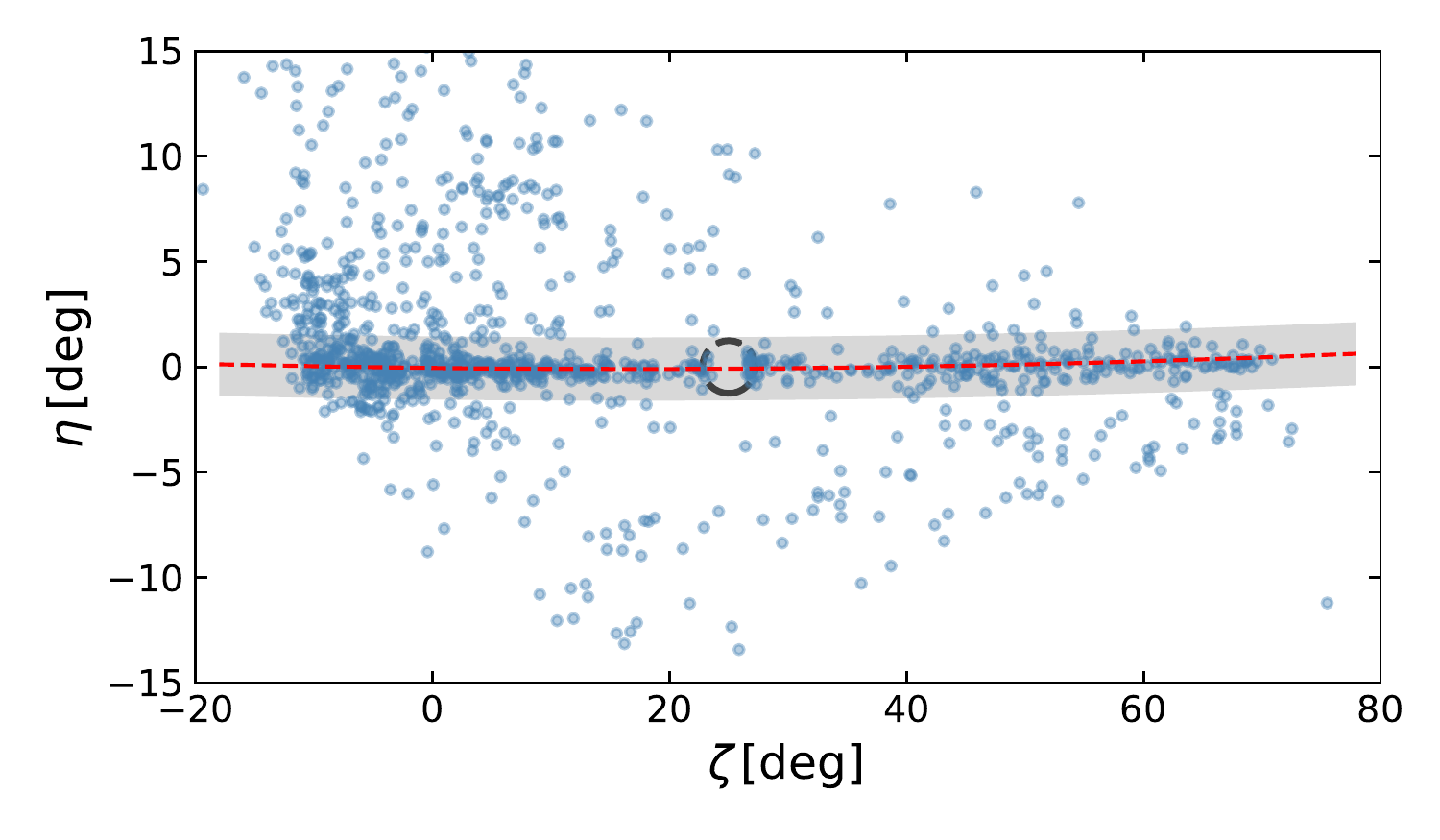}
\includegraphics[width=0.9\hsize]{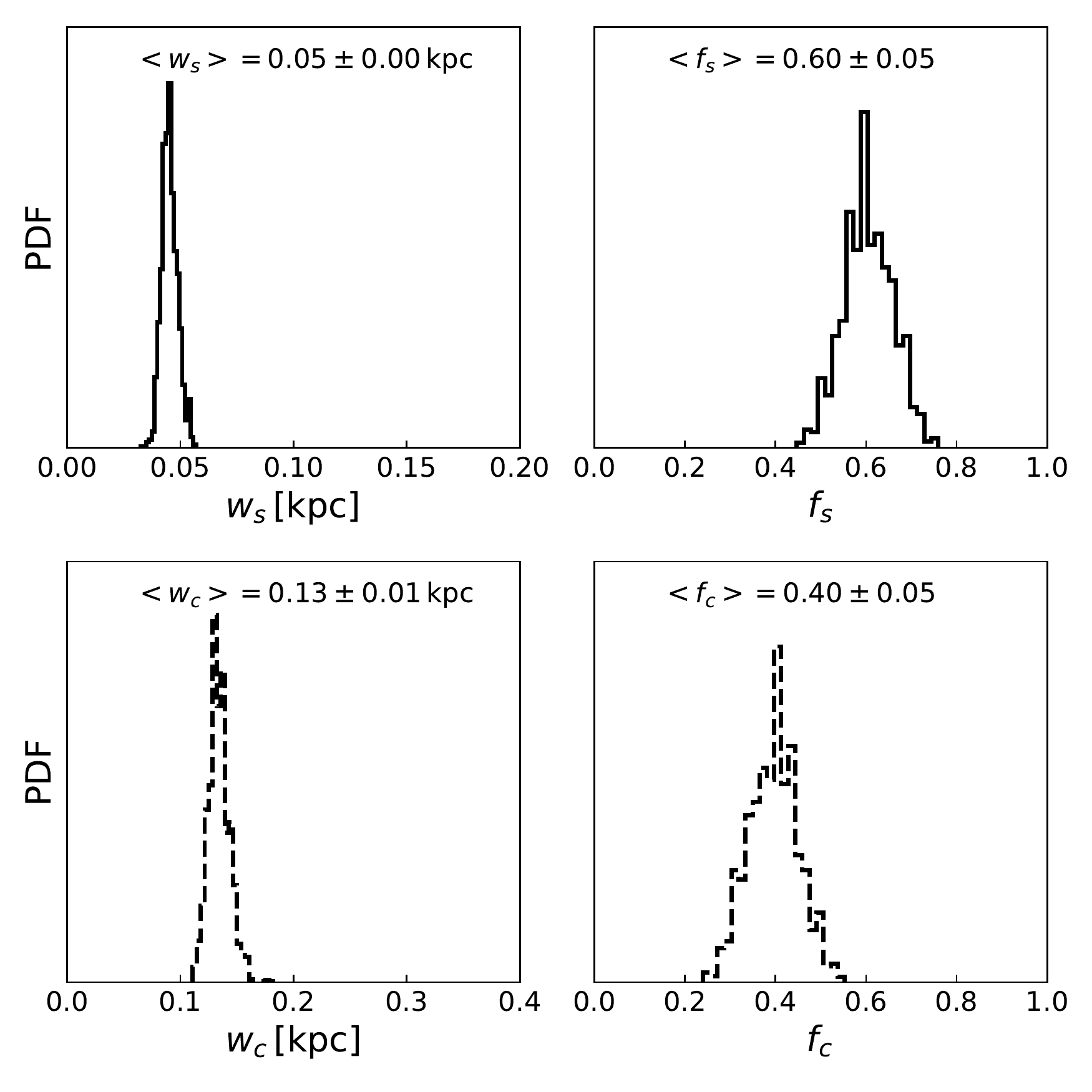}
\vspace{-0.5cm}
\end{center}
\caption{Analyzing the {\it cocoon} feature in a synthetic stream. {\it Top panel:} Distribution of stellar particles of a synthetic stream that possessed similar properties to GD-1. The circle marks the location of the progenitor that was removed intentionally. The gray band effectively corresponds to a $5\sigma$ selection width around GD-1, and we use stars within this region in our analysis to make a meaningful comparison to GD-1. {\it Bottom panels:} Probability distribution function for the Gaussian model. As expected, we find a bi-modal distribution of stellar components in this case,  corresponding to the presence of a thin structure and an accompanying extended component. The red curve in the top panel corresponds to the best fitted polynomial function using Equation~\ref{eq:polynomial}.}
\label{fig:Fig_9}
\end{figure}
\section{Interpreting the `Cocoon' with Cosmological Simulations}\label{sec:Simulations}

In order to interpret our findings, we turn to a set of cosmological simulations. In this section, we first describe the simulation setup, followed by the interpretation of the identified {\it cocoon} phenomenon.

The simulation used here to illustrate the formation of stream {\it cocoons} is essentially the same as that developed in \cite{Carlberg2018-StreamSimulation}. The simulation started with the Via Lactea II halo catalog at redshift 3. The halos were reconstituted as Hernquist spheres \citep{Hernquist1990} using dark matter particles of mass $2\times10^4\msun$. The $30$ heaviest halos $(\gtrsim 4 \times 10^8\msun)$ were provided with the tidally limited King model star clusters \citep{King1966} of mass $10^5\msun$ that were inserted in randomly oriented planes into these halos. The star particles had mass $5\msun$. The dark matter particles had a softening of $200\pc$ and the star particles had a softening of $2\pc$. The mixture of dark matter and star particles were evolved with a modified version of Gadget-2 code \citep{Springel_Gadget2_2005} that provided the expected level of two-body relaxation in the star clusters. There typically are about $300,000$ time steps for a simulation run from near redshift 3, an initial age of $2.07\Gyr$, to a current epoch age of $13.4\Gyr$.

All the simulated star particles were then transformed from the Galactocentric Cartesian frame to the Heliocentric (observable) frame using the Sun’s parameters as $R_{\odot}=8.122 \kpc$ \citep{GravityCollab2018} and $V_{\odot} = (11.1, 255.2, 7.25)\kms$ \citep{Reid2014_Sun, Schornich2010_Sun}. We then selected out only those stars that lay in the region of the sky where $|b|>20\deg$ and $5\kpc < d_{\odot} < 20\kpc$. This choice of location was made so as to focus on those streams that lie in a similar region of the sky as GD-1, thereby allowing a meaningful comparison. A few of the selected structures ($\approx 350$ structures were reviewed) are shown in Figure \ref{fig:Fig_7}. In the corresponding figure, each panel (shown in the coordinate system that roughly aligns along the streams) shows only those star particles that came from the same GC progenitor. One can see that all of these streams contain a secondary diffuse stellar component (the {\it cocoon}), similar to one that we detect here for the GD-1 stream. But how does this secondary feature actually form? The origin of this phenomenon becomes clear by examining the evolution of GCs in the simulation. 

{\it Process of cocoon formation:} In the simulation, all the stars start in GCs. These GCs are placed into a disk-like distribution in the dark matter sub-halos \citep{Mayer2001} on nearly circular orbits with radii of $\sim 1\kpc$. As the tidal fields of these sub-halos begin to strip off the stars from the GCs a ``donut'' of stars is formed, which is initially dispersed around the $\sim 1\kpc$ orbit (see Figure~\ref{fig:Fig_8}). Once the sub-halo falls into the main halo, its dark matter merges with the main halo (partially or completely) and the GC, along with the dispersed stars, gets deposited into the main halo. The GC now releases stars on its new orbit in the main halo, forming the thin and dense part of the stream, while the stars that were spread out in the ``donut'' like shape end up creating a broader stream enveloping the thin stream structure. Both the thin stream and the {\it cocoon} finally end up on quite similar orbits. These different phases of GC's dynamical evolution are illustrated in Figure~\ref{fig:Fig_8}.
 
This picture may now provide an explanation of the newly identified {\it cocoon} component we have found in this paper. According to this framework,  GD-1's progenitor GC originally formed inside a ``dwarf galaxy'' dark matter sub-halo, and was brought into the Milky Way during the accretion of the parent sub-halo. Our qualitative study of synthetic streams further suggests that the accompanying {\it cocoon} structure is most likely a ubiquitous characteristic that is exhibited by all the streams that are remnants of the accreted GCs that came along within their satellite galaxies. 

We also carried out a quantitative analysis with one of the synthetic streams to test if they too reveal a bi-modal distribution, similar to that shown above for GD-1. To this end, we first selected a candidate structure from our set of synthetic streams that shared similar structural and orbital properties to those of GD-1. We chose the stream shown in Figure \ref{fig:Fig_7}b, as it possessed similar physical length, spanning distance and $L_z$ value as that of GD-1. The stream is also shown in Figure \ref{fig:Fig_9}. Once again, we fit the same double Gaussian model via an MCMC process. So as to make a meaningful comparison with the GD-1 case, here we analyzed only those star particles that roughly lay within the physical width that was effectively equivalent to $5\sigma$ selection width of the GD-1 stars. Also, note that in this case we lack a corresponding orbital model for the stream (that basically goes into the equation \ref{eq:test1} in the form of the parameter ${\rm \phi^m_2}$). For this, we allow ${\rm \phi^m_2}$ ($\eta^m$ in the present case) to be an additional parameter of our model that is fitted to the data using the polynomial parameterization of the form:
\begin{equation}\label{eq:polynomial}
\begin{split}
\eta^m = a + b\zeta + c\zeta^2 \, ,
\end{split}
\end{equation}
where $a, b, c$ are the intrinsic parameters of $\eta^m$ that are actually sampled during the MCMC exploration. The physical dispersions of the two structural components (the stream and the {\it cocoon}) are then calculated about the fitted $\eta^m$. The results obtained in this case are shown in Figure~\ref{fig:Fig_9}. As expected, we find a bi-modal distribution of the particles corresponding to a thin structure and an accompanying diffuse component. In this case, the secondary component was identified with a physical width of $\langle {w_c} \rangle =130\pm10\pc$ (comparable to the size of GD-1's {\it cocoon}, however note that the structure actually extends well beyond this range, similar to the range displayed in Figure~\ref{fig:Fig_1}b). The similarity between the results obtained in this case, by analyzing synthetic stream structures from the cosmological simulations, with the ones obtained for the GD-1 case, that was based on astrophysical data, establishes both the plausibility of the existence and the positive detection of the {\it cocoon} structure around the GD-1 stream.

\begin{figure}
\begin{center}
\includegraphics[width=0.9\hsize]{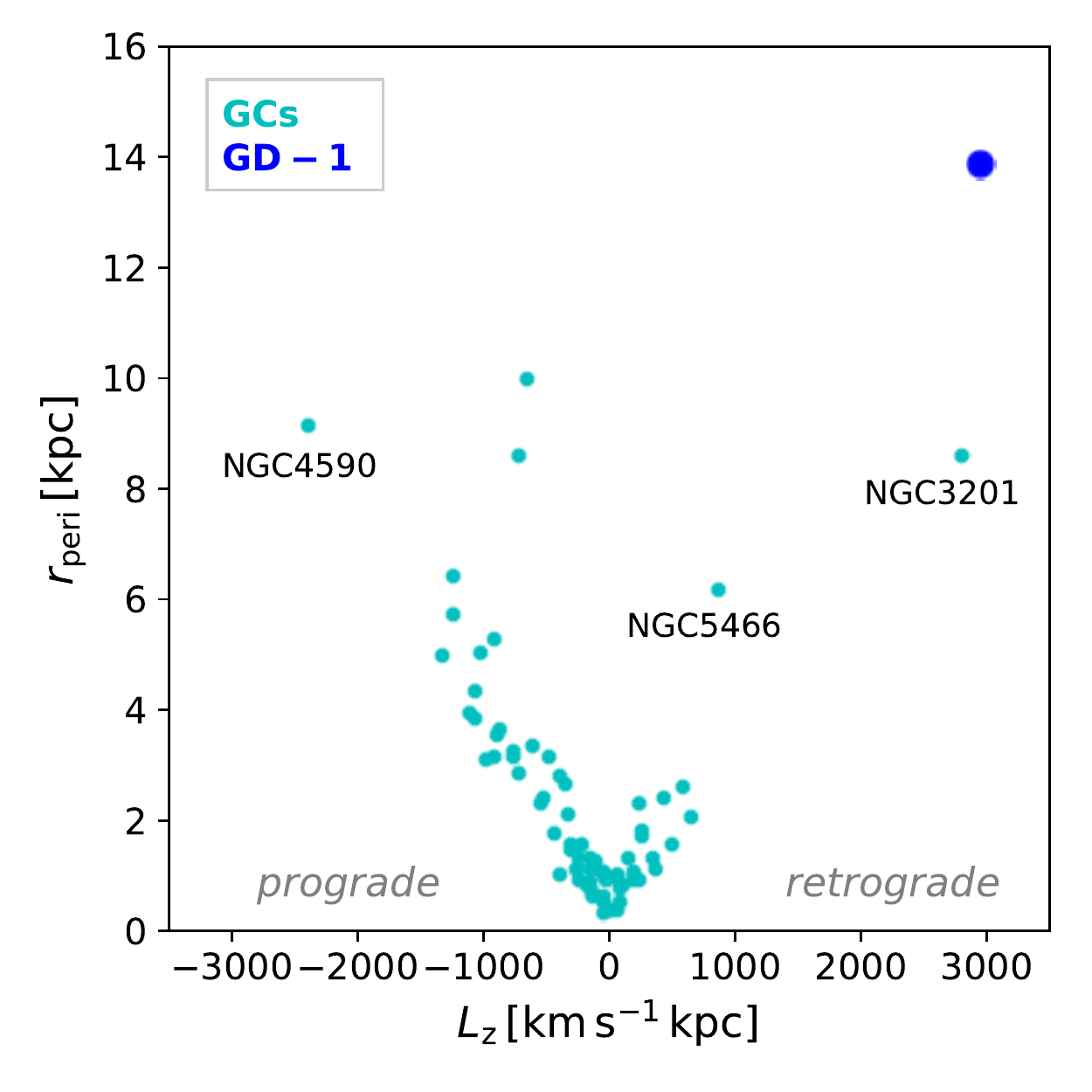}
\end{center}
\vspace{-0.5cm}
\caption{Orbital properties of GD-1 in comparison to Milky Way GCs. GCs with orbits measured by Gaia \citep{GaiaCollab2018kinematics} are shown in the $L_{\rm z}$ -- $r_{\rm peri}$ plane, along with the GD-1 stream. Some well-known GCs have been labelled. In this plane, objects with retrograde motion have positive $L_{\rm z}$. The orbital properties of GD-1 are clearly extreme. The highly retrograde orbit suggests an accretion origin for the stream progenitor, which is consistent with the scenario presented here.}
\label{fig:Fig_Lz}
\end{figure}
\section{Conclusion and Discussion}\label{sec:Discussion}

We investigated the structure and morphology of the GD-1 stellar stream in order to understand its embryonic origin and its participation in the formation of the Milky Way halo. Being a remnant of a GC, GD-1 is often conceived as a simple system that is usually approximated as a linear structure. Our study here suggests otherwise. 

Using the 6D phase-space and photometric measurements, we probed the sky region around the GD-1 stream and found evidence for the presence of an additional extended and diffuse stellar component (which can be perceived in Figure~\ref{fig:Fig_1}b). This secondary component, that we dub the {\it cocoon} here, was detected at a $>5\sigma$ confidence level. The physical width of the {\it cocoon} was found to be $\langle {w_c} \rangle =110\pm20\pc$ (Figures~\ref{fig:Fig_4},\ref{fig:Fig_5}), which we obtained by analyzing the 6D phase-space and color-magnitude region within the $5\sigma$ volume around the GD-1 stream. Alongside, the revised physical width of the narrow component was estimated to be $\langle {w_s} \rangle \sim 30\pc$. To interpret this detection, we turned to a set of cosmological simulations, and found that this {\it cocoon} feature is most likely a ubiquitous characteristic that is exhibited by all the streams that are remnants of the accreted GCs that came within dwarf galaxy-mass dark matter sub-halos (Figure~\ref{fig:Fig_7} and \ref{fig:Fig_8}, see section~\ref{sec:Simulations} for details on how the {\it cocoon} structure is formed). This framework, in light of the detection of GD-1's {\it cocoon}, lends credence to the idea that GD-1's progenitor GC was originally brought into the Milky Way in a now defunct satellite galaxy. 

The highly retrograde nature of GD-1's orbit ($L_{\rm z}\sim3000 \kms\kpc$, see Figure \ref{fig:Fig_Lz}, \citealt{Malhan2018PotentialGD1}) together with its metal poor stellar population paints a similar picture that appears consistent with this scenario. This is because more retrograde motions and lower metallicties in the outer galactic halo are indicative of accretion of low-mass galaxies \citep{Carollo2007metallicityhalo}. Moreover, the detection of the {\it cocoon} means that GD-1's progenitor was still within the ``dwarf galaxy'' dark halo prior to the merging of the dark matter sub-halo onto the main halo.

In principle, there are other physical processes that can also give rise to similar extended and diffuse components in (otherwise) thin stellar streams. The thick component could be produced by the perturbation effects of the stream's interaction with other massive galactic components, such as by shocks from the disk. However, GD-1's disk crossings take place between $13\kpc$ and $23\kpc$ from the Galactic center, where the disk density is too low to significantly impact the stream \citep{Bonaca_spur_2018}. Interactions with dark matter sub-halos could also result in the heating of stellar streams \citep{Ibata_2002DM_TS, Johnston2002DM_TS, Gaskins2008_streamGaps, StreamGap_Carlberg2012, StreamGap_Erkal2016}. However, GD-1's velocity dispersion is as low as $\sim 1.5\kms$ \citep{Malhan2018PotentialGD1}, indicating dynamical coldness of this system, that suggests that so far GD-1 has not suffered substantial external heating. Recently \cite{Carlberg2018Density_Structure} argued that another contribution to stream density variations could also stem from the GC streams having traversed the large-scale tidal field of the host galaxy which varied over time as the galaxy assembled.

The cosmological simulations suggest that we have possibly found an unambiguous means of distinguishing between the {\it in-situ} and {\it ex-situ} formed GC streams population, which otherwise can be hard to disentangle. The number of streams identified in this manner should in principle place a lower limit on the number of past accretion events, allowing one to quantify the number of stars in the stellar halo that are a result of hierarchical merging events \citep{Bullock_n_Johnston2005}. Although here we studied only a single GC stream, making the case for the discovery of the {\it cocoon} phenomenon, it would be interesting to analyze other GC streams of the Milky Way in order to examine if the {\it cocoon} property is ubiquitous, or is limited only to particular types of globular cluster streams. 
    
The simulations that we studied here show a large diversity in the structural morphology of the {\it cocoon} component (Figure~\ref{fig:Fig_7}), nevertheless, the phenomenon was found to be ubiquitous among all the accreted GC streams. The morphology would of course depend on the orbital history of the accreted satellite, but it may retain information about its now defunct dark nursery. Through careful N-body modeling of the GD-1 stream, which can simultaneously reproduce its {\it cocoon}, it may be possible to constrain the initial conditions of the parent dark matter sub-halo. For instance, the {\it cocoon}'s phase-space density may be linked with the dark matter density profile and its phase-space dispersion with the mass and the physical size of the dark sub-halo.  For example, it is known that for the same set of initial conditions the GC stripping occurring in a ``cuspy'' dark matter profile is relatively more pronounced than in a ``cored'' profile (since force fields in constant density profiles are rather compressive in nature, \citealt{Cole2012}, \citealt{Petts2016_dynamical_fric}, \citealt{Contenta2018EridanusII}). In such a case, ``cuspy'' profiles are expected to form denser cocoon, which can then ultimately get reflected in the morphology of the accreted stream and {\it cocoon} system. This would ofcourse also be sensitive to the initial phase-space position of the GC within the dark sub-halo. Such a study, employing the {\it cocoon} as a dark matter probe, may open an exotic window on the pre-merging times by revealing the physical properties of the primordial dark sub-halos. The inferred properties, in principle, could be different from those that are currently observed for the dwarf galaxies. Such a comparison would also be useful in understanding and testing the galaxy evolution paradigms and cosmological models for the lowest mass halos.


\section*{ACKNOWLEDGEMENTS}
We thank the anonymous referee very much for their helpful comments.

K.M. and K.F. acknowledge  support  by the $\rm{Vetenskapsr\mathring{a}de}$t (Swedish Research Council) through contract No.  638-2013-8993 and the Oskar Klein Centre for Cosmoparticle Physics.  K.F. acknowledges support from DoE grant de-sc0007859 at the University of Michigan as well as support from the Leinweber Center for Theoretical Physics. M.V. is supported by NASA-ATP award NNX15AK79G.

This work has made use of data from the European Space Agency (ESA) mission {\it Gaia} (\url{https://www.cosmos.esa.int/gaia}), processed by the {\it Gaia} Data Processing and Analysis Consortium (DPAC, \url{https://www.cosmos.esa.int/web/gaia/dpac/consortium}). Funding for the DPAC has been provided by national institutions, in particular the institutions participating in the {\it Gaia} Multilateral Agreement. 

Guoshoujing Telescope (the Large Sky Area Multi-Object Fiber Spectroscopic Telescope LAMOST) is a National Major Scientific Project built by the Chinese Academy of Sciences. Funding for the project has been provided by the National Development and Reform Commission. LAMOST is operated and managed by the National Astronomical Observatories, Chinese Academy of Sciences. 

Funding for SDSS-III has been provided by the Alfred P. Sloan Foundation, the Participating Institutions, the National Science Foundation, and the U.S. Department of Energy Office of Science. The SDSS-III web site is \url{http://www.sdss3.org/}.

SDSS-III is managed by the Astrophysical Research Consortium for the Participating Institutions of the SDSS-III Collaboration including the University of Arizona, the Brazilian Participation Group, Brookhaven National Laboratory, Carnegie Mellon University, University of Florida, the French Participation Group, the German Participation Group, Harvard University, the Instituto de Astrofisica de Canarias, the Michigan State/Notre Dame/JINA Participation Group, Johns Hopkins University, Lawrence Berkeley National Laboratory, Max Planck Institute for Astrophysics, Max Planck Institute for Extraterrestrial Physics, New Mexico State University, New York University, Ohio State University, Pennsylvania State University, University of Portsmouth, Princeton University, the Spanish Participation Group, University of Tokyo, University of Utah, Vanderbilt University, University of Virginia, University of Washington, and Yale University. 

\bibliographystyle{apj}
\bibliography{ref1}

\end{document}